\documentclass[aps,pre,preprint, showpacs, superscriptaddress]{revtex4}
\usepackage{latexsym}
\usepackage{natbib}
\usepackage{amsmath}
\usepackage{amsfonts}
\usepackage{amssymb}
\usepackage[T1]{fontenc}
\usepackage[dvips]{color,graphicx}   
\usepackage{verbatim}   
\usepackage{t1enc}
\usepackage{anysize}
\usepackage{tocbibind}
\usepackage{subfig}
\usepackage[percent]{overpic} 
\usepackage{float}  
\usepackage{appendix} 
\usepackage{booktabs} 
\usepackage{natbib}
\usepackage{multirow}
\usepackage{array}

\begin{document}
\title{Triple point induced by targeted autonomization on interdependent networks}
\author{L. D. Valdez} \affiliation{Instituto de Investigaciones
  F\'isicas de Mar del Plata (IFIMAR)-Departamento de F\'isica,
  Facultad de Ciencias Exactas y Naturales, Universidad Nacional de
  Mar del Plata-CONICET, Funes 3350, (7600) Mar del Plata, Argentina.}
  \author{P. A. Macri} \affiliation{Instituto de Investigaciones
  F\'isicas de Mar del Plata (IFIMAR)-Departamento de F\'isica,
  Facultad de Ciencias Exactas y Naturales, Universidad Nacional de
  Mar del Plata-CONICET, Funes 3350, (7600) Mar del Plata, Argentina.}
  \author{L. A. Braunstein} \affiliation{Instituto de Investigaciones
  F\'isicas de Mar del Plata (IFIMAR)-Departamento de F\'isica,
  Facultad de Ciencias Exactas y Naturales, Universidad Nacional de
  Mar del Plata-CONICET, Funes 3350, (7600) Mar del Plata, Argentina.}
  \affiliation{Center for Polymer Studies, Boston University, Boston,
  Massachusetts 02215, USA}

\begin {abstract}
Recent studies have shown that in interdependent networks an initial 
failure of a fraction $1-p$ of nodes in one network,
exposes the system to cascade of failures. Therefore it is important to
develop efficient strategies to avoid their collapse. Here, we
provide an exact theoretical approach to study the evolution of the
cascade of failures on interdependent networks when a fraction
$\alpha$ of the nodes with higher connectivity are autonomous. We
found, for a pair of heterogeneous networks, two critical percolation
thresholds that depend on $\alpha$, separating three regimes with very
different network's final sizes that converge into a triple point in
the plane $p-\alpha$. Our findings suggest that the heterogeneity of
the networks represented by high degree nodes is the responsible of
the rich phase diagrams found in this and other investigations.
\end{abstract}


\maketitle

\section{Introduction}
Networks of networks are systems composed by several networks that in
many cases depend on each other in a nontrivial
way~\cite{Rin_01,Kiv_01}. An example of such systems are the power
grid and the communication networks \cite{Ros_01} in which the first
one provides electric power to the communication network, and the last
one provides control service to the electric network. Another example
is the traffic flow between cities, through the sea port and airport
networks~\cite{Ron_02} in which the flow of individuals or goods in a
city decays if it does not receive traffic from one of these
networks. Most of these systems are composed by individual networks
connected by internal connectivity links. The role of these internal
links is to generate a single component network that allows to
distribute some entities between the nodes, such as the electric flow in
the power grid network. Nodes of different networks are connected by
interdependent links that enable the support between them.  Thus in
general, when a node fails in one network, the failure propagates to
the other networks through the interdependent links producing
sometimes a ``domino'' effect with harmful consequences for the
functionality of the networks.

It was shown that under a failure of a fraction of nodes in one
network, the interdependency can produce a cascade of failures that
spreads through all the system with catastrophic consequences in the
robustness of the individual networks.  Buldyrev $et\;al.$
\cite{Bul_02} proposed a minimalist model, based on percolation
theory, to study the dynamics of the cascade of failures.  In
Ref.~\cite{Bul_02} they consider two interdependent networks, denoted
by $A$ and $B$ with fully interdependency $i.e.$ each node depends on
a node in the other network. By definition, a functional node is
connected to the giant component (GC) of its own network and depends
on a node in the other network that also belongs to its GC. Otherwise
the node is dysfunctional, $i.e.$ it is failed~\cite{Bul_02}. Thus,
the GC is the only ``functional cluster'' and there is only one in
each network. Before receiving an initial failure or attack, every
node in one network is supported by its interdependent node in the
other network and thus the nodes are fully interdependent with a full
correspondence between the sizes of the GC of both networks.  The
random failure of an initial fraction $1-p$ of nodes in one network
triggers the cascade of failures and, as a consequence, the
correspondence between both GC is broken.  At each time step, the
dysfunctional nodes transmit the failure to their interdependent
neighbors, producing dysfunctional nodes in the other network.  The
process reaches the steady state when both networks are abruptly
destroyed with a first order transition at a critical threshold $p_c$
or when above $p_c$ the correspondence between the GC is
reestablished. It was shown that fully interdependent networks are
very fragile under random failures, $i.e$, they have a higher critical
threshold $p_c$ than isolated networks, regardless of the degree
distribution~\cite{Bul_02}. This was an exciting result because it is
well known that isolated heterogeneous scale-free (SF) networks are
very robust against random failures ($p_c\to 0$). However, to consider
full interdependency is not very realistic because nodes in each
network can work autonomously. For example, some nodes can have a
``power supply'' or a backup that allow them to remain functional even
when they lack of support from the other network, increasing their
chance to remain functional. As a consequence, partial interdependency
where a random fraction $q < 1$ of nodes are interdependent and the
rest are autonomous, increases the robustness of the individuals
networks compared to the case of fully interdependency~\cite{Ron_01,
  Di_01, Gao_03, Gao_01, Gao_02}. It was found that depending on the
value of $q$ and on the fraction $1-p$ of random failures in network
$A$, the transition changes from a discontinuous to a continuous one.
In these partial interdependent networks the correspondence in the
steady state between networks $A$ and $B$ is broken because the
autonomous nodes in network $B$ do not receive the initial failure of
network $A$ and can only become dysfunctional by the failure of
non-autonomous nodes that disconnect them from its GC. Then, in the
steady state, the size of the GC of network $B$ is bigger than the one
in network $A$.  In Ref.~\cite{Di_01} it was found that for
heterogeneous SF under random autonomization the sizes of the
functional clusters undergo an abrupt decreasing for a certain value
of $p=p_{c}^{+}$ without a full collapse due to the fact that the high
degree nodes are sustained by autonomous nodes~\cite{Exxon_02}.
However bellow $p_{c}^{+}$ the size of the GC of network $A$ decreases
continuously to zero as in a second order percolating transition at a
value $p=p_{c}^{-}$, while the size of network $B$ goes to a finite
value.  The goal is to find a way to autonomize efficiently the
networks in order to increase their robustness compared to the case of
random autonomization~\cite{Sch_01,Yag_01}.  Schneider
$et\;al.$ \cite{Sch_01} proposed a model where the robustness of the
system is enhanced by targeted autonomization of a fraction $\alpha
\equiv 1-q$ of the higher degree nodes. Using a theoretical mean field
approximation that assumes that the cascade of failures affects both
autonomous and non-autonomous nodes, they showed that even for
homogeneous networks there is a critical point in the plane $p-q$, at
$(p_c,q_c)$ where the transition changes from first order for $q>q_c$
to a continuous one for $q<q_c$. This theoretical results was
qualitatively supported by simulation, but the exact theoretical
solution was not derived so far. An exact theoretical formulation
allows to find some other effects that are hidden in the simulations
due to finite size effects. Very recently, Valdez
$et\;al.$~\cite{Val_01}, introduce an exact general framework that they
apply to explain the effect of partially correlated interdependent
networks in the robustness of heterogeneous SF interdependent networks
under cascade of failures. The exact result allowed to find very
interesting features such as a triple point in the phase diagram that
depend on the level of correlation. Here, we apply the formalism
presented in Ref.~\cite{Val_01} to targeted autonomization and derive
the exact theoretical solutions for this process.

\section{Theoretical results}
We study the temporal evolution of the sizes of the GC of two
interdependent networks under targeted autonomization when a fraction
$\alpha$ of the higher degree nodes of both networks are autonomous.
Each network, that we denote by $A$ and $B$, has connectivity links
distributed according to $P[k_{A}]$ and $P[k_B]$, where $k_A$ and
$k_B$ are the connectivity links of nodes in $A$ and $B$ respectively.
Let's assume that a fraction of interdependent nodes $q_{A}[k_A]$
($q_B[k_B]$) in network $A$ ($B$) depends on the connectivity links of
network $A$ ($B$).  In the initial stage a fraction $1-p$ of nodes
fails at random in network $A$.  At each stage $n$ of the cascade
failure that goes from $A$ to $B$, a node in network $A$ with degree
$k_A$ is functional if it is autonomous and belongs to its GC with
probability $(1-q_{A}[k_A]) (1- (1-p\;f_{An})^{k_A})$ or if it is not
autonomous but was connected to the GC of $B$ in a previous stage with
probability $q_{A}[k_A]\bigl(1-(1-f_{Bn-1})^{k_B}\bigr)$. Since the
initial failure (at $n=0$) of $1-p$ nodes happens only in network $A$,
then only $f_{An}$ is multiplied by $p$.  Here $f_{An}$ ($f_{Bn}$) is
the probability that a random selected edge that leads to a non-failed
node at $n=0$, this node belongs to the GC of network $A$ ($B$) at
stage $n$~\cite{Bra_01, New_01} and fulfills the self consistent
equation
\begin{eqnarray}\label{eq2}
f_{An}&=&\sum_{k_A=k_{min}}^{k_{max}} \frac{k_A \;P[k_A]}{\langle k_A
  \rangle}\left(1-q_{A}[k_A]\right)(1-(1-pf_{An})^{k_A-1} )
+\nonumber\\&&\sum_{k_A=k_{min}}^{k_{max}} \frac{k_A
  P[k_{A}]}{\langle k_A \rangle} q_{A}[k_A](1-(1-pf_{An})^{k_A-1})\times\nonumber\\&&
\sum_{k_B=k_{min}}^{k_{max}} P[k_{B}]
(1-(1-f_{Bn-1})^{k_B}).
\end{eqnarray}
where $k_{min}$ and $k_{max}$ are the minimum and maximum
connectivity links respectively.

The first term in Eq.~(\ref{eq2}) takes into account the autonomous
functional nodes in $A$ with degree $k_A$ and the second term
corresponds to functional nodes in $A$ with degree $k_A$ that depend
on functional nodes of $B$ with degree $k_{B}$ at step $n-1$. 
Thus, the fraction of nodes $\Psi_{n}$ of the GC of network $A$ at
step $n$ is given by,
\begin{eqnarray}\label{eq1}
\Psi_{n}&=&p\Biggl(\sum_{k_A=k_{min}}^{k_{max}} P[k_A](1-q_{A}[k_A])
(1-(1-pf_{An})^{k_A})+\nonumber\\&&\sum_{k_A=k_{min}}^{k_{max}} \;
q_A[k_A] P[k_{A}] (1-(1-pf_{An})^{k_A})\times\nonumber\\&& \sum_{k_B=k_{min}}^{k_{max}}
P[k_{B}] (1-(1-f_{Bn-1})^{k_B})\Biggr),
\end{eqnarray}

For network $B$, $f_{Bn}$ also fulfills a self-consistent equation
\begin{eqnarray}\label{eq4}
&f_{Bn}&=\sum_{k_B=k_{min}}^{k_{max}}\frac{k_BP[k_B]}{\langle k_B \rangle}(1-q_{B}[k_B])(1-(1-f_{Bn})^{k_B-1})+\nonumber\\&&p\sum_{k_B=k_{min}}^{k_{max}}\frac{k_BP[k_B]}{\langle k_B \rangle}q_{B}[k_B](1-(1-f_{Bn})^{k_B-1})\times\nonumber\\&&\sum_{k_A=k_{min}}^{k_{max}}P[k_A](1-(1-pf_{An})^{k_A}).
\end{eqnarray}

Thus the fraction of nodes $\phi_{n}$ of the GC of network $B$ is given by
\begin{eqnarray}\label{eq3}
\phi_{n}&=&\sum_{k_A=k_{min}}^{k_{max}} P[k_B](1-q_{B}[k_B])(1-(1-f_{Bn})^{k_B})+\nonumber\\&&p\sum_{k_A=k_{min}}^{k_{max}} P[k_{A}] (1-(1-pf_{An})^{k_A})
\times\nonumber\\&&\sum_{k_B=k_{min}}^{k_{max}} q[k_{B}] P[k_{B}] (1-(1-f_{Bn})^{k_B}),
\end{eqnarray}
In the steady state, $i.e.$ for $n\to \infty$, $\Psi_{n}\approx
\Psi_{n-1}$ and $\phi_{n}\approx \phi_{n-1}$, thus $\Psi_{n}$ and
$\phi_{n}$ converges to $\Psi_{\infty}$ and $\phi_{\infty}$,
respectively ~\cite{Son_01,Bax_01,Val_01}.

In this model, $q_{A}[k_A]$ and $q_{B}[k_B]$ are given by
\\
\[
q_{i}[k_i] = \left\{%
\begin{array}{ll}
1 &  k_i < k_S\\
(1-w)  & k_i=k_S\\
0  & k_{S} < k_i,
\end{array}%
\right.
\]
\\
where $i=A, B$, $k_S$ is the degree at and above which a fraction
$\alpha$ of nodes are autonomous, and $k_S$ fulfills $
\sum_{k_i=kmin}^{k_{s}-1}P[k_i]\leq 1-\alpha
<\sum_{k_i=kmin}^{k_{s}}P[k_i] $. Thus if we denote by $w$ the
fraction of autonomous nodes with degree $k_s$,
$wP[k_s]+\sum_{k_i=k_{s}+1}^{k_{max}}P[k_i]=\alpha$.  Note that for
$\alpha>0$ Eqs.~(\ref{eq2})-(\ref{eq3}) are not symmetric which leads
to the non correspondence between the final sizes of the GC of
networks $A$ and $B$, as mentioned above. The symmetry is restored
only for $q_{A}=q_{B}=1$ ($\alpha=0$)~\cite{Bul_02} or when the
initial failure happens in both networks~\cite{Di_01}.

\begin{figure}[H]
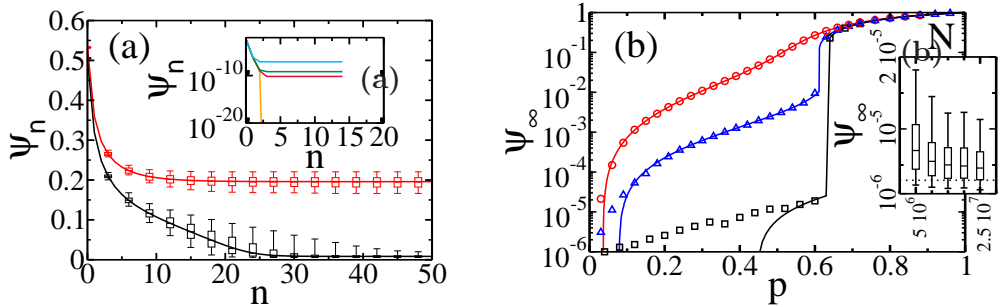

\centering
\vspace{1.0cm}
  \begin{overpic}[scale=0.21]{Fig1a.eps}
    \put(80,50){\bf{(a)}}
 \end{overpic}\hspace{0.25cm}
  \hspace{0.25cm}
  \begin{overpic}[scale=0.21]{Fig1b.eps}
    \put(80,50){{\bf{(b)}}}
  \end{overpic}\hspace{0.25cm}
  \hspace{0.25cm}
\caption{(Color on-line). Cascade failure on network $A$ with targeted
  autonomization on SF networks of size $N=10^6$ with $\lambda=2.5$
  and $2 \leq k \leq k_{max}=1000$.  In Fig.~(a) we show $\Psi_{n}$ as
  a function of $n$ obtained by 100 network realizations (box plots)
  and from Eqs.~(\ref{eq2})-(\ref{eq3}) (solid lines) for
  $\alpha=0.01\%$, $p=0.60$ (black) and $p=0.62$ (red). The ends of
  the whiskers represent the $5th$-percentile and
  $95th$-percentile. In the inset we plot the theoretical solution for
  (from top to bottom): $p=0.07790$, $p=0.077822$, $p=0.0778214$ and
  $p=0.0778213$ ($p_{c}\approx 0.077821334$). In Fig.~(b) we plot
  $\Psi_{\infty}$ as a function of $p$ for different values of
  $\alpha$, $\alpha=0.1\%$ (red, $\bigcirc$), $\alpha=0.01\%$ (blue,
  $\triangle$) and $\alpha=0.001\%$ (black, $\square$). The
  simulations are represented by symbols and the theoretical solutions
  obtained from Eqs.~(\ref{eq2})-(\ref{eq3}) for $n\to \infty$ by
  solid lines. The simulations were performed over 100 network
  realizations. In the inset we show the finite size effects of the
  simulations as $N$ increases from $N=5\;10^{6}$ to $N=2.5\;10^{7}$
  for $p=0.47$ and $\alpha=0.001\%$. Each box plot is obtained over
  $1000$ network realizations and shows the $5th$, $25th$, $50th$,
  $75th$ and $95th$ percentile values. For bigger network sizes, the
  median of $\Psi_{\infty}$ approachs to the theoretical value (dotted
  line).}\label{fig_target}
\end{figure}
We apply our equations to SF networks of sizes $N=10^6$ with degree
distribution $P[k]\sim k^{-\lambda}$ with $k_{min}\leq k \leq k_{max}$
and $\lambda=2.5$. Here, we use $k_{min}=2$ to ensure that at the
beginning all the nodes belong to the GC~\cite{Bor_01}. Since the
theoretical solutions of Eqs.~(\ref{eq2})-(\ref{eq3}) near the
criticality are sensitive to the precision employed in the
calculations, we use a multiple precision arithmetic
library~\cite{Nik_01}. We chose a finite $k_{max}$ in order to emulate
the finite power law region observed in many real
networks~\cite{Ama_01} such as the movie actor network~\cite{Bar_02},
the scientific collaboration network~\cite{Bar_01,New_01} and the
protein network~\cite{Jeo_01}. In Fig.~\ref{fig_target} we show the
temporal evolution and the steady state of the fraction of nodes in
the GC of network $A$, for SF networks with $\lambda=2.5$. From
Fig.~\ref{fig_target} we can see the excellent agreement between the
theory and the simulations, except for very low values of
$\Psi_{\infty}$ that is due to finite size effects. In the inset of
Fig.1b we show from $N=5\;10^6$ to $N=2.5\;10^7$ that the value of
$\Psi_{\infty}$ obtained from the simulations approaches to the
theoretical solution for $p=0.47$ and $\alpha=0.001\%$ as the system
size increases.

\begin{figure}[H]
\centering
\vspace{1cm}
 \includegraphics[scale=0.30]{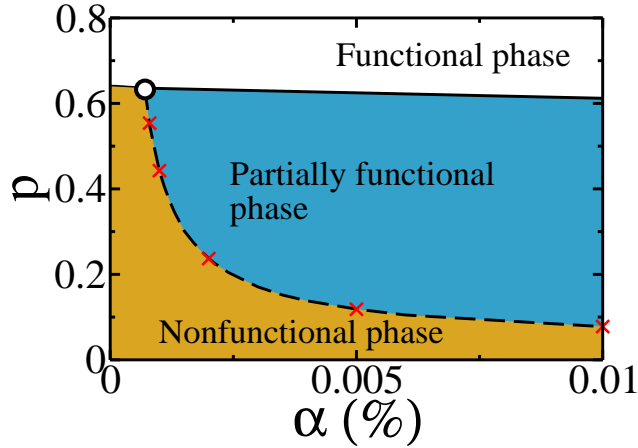}
  \vspace{0.5cm}
\caption{(Color on-line) Phase diagram in the plane $\alpha-p$: i) the
  yellow area corresponds to the nonfunctional phase, $i.e$,
  $\Psi_{\infty}=0$, ii) the blue area corresponds to a partial
  functional phase in which the size of the GC of both networks is
  $\lesssim 10^{-3}$ and iii) the white area corresponds to a
  functional phase where $\Psi_{\infty}\gtrsim 10^{-2}$. The white
  circle corresponds to a triple point. The solid lines represent the
  abrupt change on both network's sizes and the dotted line, which is
  defined for $\alpha>\alpha_{c}$, represents a continuous transition
  of $\Psi_{\infty}$ at $p_{c}^{-}$, obtained from
  Eq.~(\ref{eq7}). The cross symbols correspond to some points
  obtained from Eqs.~(\ref{eq2})-(\ref{eq3}) around which the
  solution $\Psi_{n}$ vanishes for $n \to \infty$.}\label{fig.phase}
\end{figure}

As was observed in Ref.~\cite{Sch_01}, for homogeneous networks, the
robustness of the networks increases with $\alpha$ due to the fact
that the higher degree nodes of both networks, that are the ones that
sustain the functionality of the networks, are autonomous. Our
theoretical equations allow to find a surprising behavior of the
transitions with two critical thresholds at $p=p_{c}^{+}$ and
$p=p_{c}^{-}$ that depend on $\alpha$ (with
$p_{c}^{-}<p_{c}^{+}$)~\cite{Exxon_03}. At $p_{c}^{+}$ the sizes of
the functional networks $A$ and $B$ have an abrupt jump. Below this
critical threshold the GC of network $A$ is destroyed at
$p_{c}^{-}$. To compute the value of the critical point $p_{c}^{+}$,
we solve numerically the system of Eqs.~(\ref{eq2}) and (\ref{eq4})
with the condition $\det(J-I)=0$ \cite{Exxon_01}, where $J$ is the
Jacobian of Eq.~(\ref{eq2}) and (\ref{eq4}) and $I$ is the identity
matrix.  This method also can be applied to find $p_{c}^{-}$, however
here we use a more explicit and physical derivation to compute
it~\footnote{Another method to obtain graphically the value of the
  critical point is by measuring the position of the peak of the NOI
  curve as a function of $p$, where the NOI is the number of
  iterations needed to reach the steady state of the evolution
  equations. At $p_{c}^{+}$ there is a sharp peak, which corresponds
  to the condition $det(J-I)=0$. However, around $p_{c}^{-}$ the NOI
  has not a visible peak within the precision we used.}. Assuming that
the transition in network $A$ is continuous, then the probability
$f_A\to 0$ continuously when $p\to p_{c}^{-}$. As a consequence at
this threshold $f_{B}$, that is not zero due to the broken symmetry
imposed by the initial failure in $A$ and by the partially
interdependency ($0 < \alpha < 1$), reduces to
\begin{eqnarray}\label{eq5}
f_{B}&=&\sum_{k_B=k_{min}}^{k_{max}}\frac{k_BP[k_B]}{\langle k_B \rangle}(1-q_{B}[k_B])(1-(1-f_{B})^{k_B-1}).
\end{eqnarray}
Solving this self consistent equation we found the non trivial
solution of Eq.~(\ref{eq5}), from where we obtain $f_B$ at the
threshold $p_{c}^{-}$.  Since $p_{c}^{-}$ is a critical point for
network $A$, the r.h.s. of Eq.~(\ref{eq2}) for $n\to \infty$ is
tangent to the identity function evaluated at $f_{A}=0$, thus
\begin{eqnarray}\label{eq6}
1&=&p\sum_{k_A=k_{min}}^{k_{max}}\frac{k_A(k_A-1)P[k_A]}{\langle k_A
  \rangle}\left(1-q_{A}[k_A]\right)+\nonumber\\&&p\sum_{k_A=k_{min}}^{k_{max}}
q_A[k_A] \frac{k_A(k_A-1)P[k_{A}]}{\langle k_A \rangle}
\sum_{k_B=k_{min}}^{k_{max}} P[k_B](1-(1-f_{B})^{k_B}).
\end{eqnarray}
Then $p=p_{c}^{-}$ is explicitly given by
\begin{eqnarray}\label{eq7}
p_{c}^{-}&=&\bigg[\frac{\langle k_{A}^2\rangle-\langle k_{A}\rangle}{\langle k_{A}\rangle}-G_{0B}[1-f_{B}]\bigg(\sum_{k_A=1}^{k_{s}-1}\frac{k_{A}(k_A-1)P[k_A]}{\langle k_{A} \rangle}+\nonumber\\&&\frac{(1-w)k_s(k_s-1)P[k_s]}{\langle k_{A} \rangle}\bigg) \bigg]^{-1},
\end{eqnarray}
where $(\langle k_{A}^2\rangle-\langle k_{A}\rangle)/\langle
k_{A}\rangle$ is the branching factor of random percolation in network
$A$ and $G_{0B}[x]\equiv \sum_{k_B}P[k_B]x^{k_B}$ is the generating
function of network $B$. Thus $p_{c}^{-}$ is a correction to the
threshold of percolation in individual networks where $p_c=\langle
k_{A} \rangle/(\langle k_{A}^2\rangle-\langle k_{A}\rangle)$, because
the branching factor in this process is reduced by the second term, as
a result of the targeted autonomization.  Note that if $k_{max}\to
\infty$ the branching factor diverges, and $p_{c}^{-}\to 0$ for all
$\alpha>0$~\cite{Exxon_04}. The solution of Eq.~(\ref{eq7}), has a
physical meaning only if $p_c^{-}<p_{c}^{+}$, otherwise there is only
one threshold at $p=p_{c}^{+}$ where both networks fully
collapses. The phase diagram in the plane $p-\alpha$, displayed in
Fig.~\ref{fig.phase}, shows a triple point in which the line of the
first order transition forks at $\alpha_c= 0.000702(1)\%$ into two
branches where the upper one corresponds to an abrupt collapses at
$p=p_{c}^{+}$ and the lower one corresponds to $p_{c}^{-}$ where the
size of network $A$ continuously vanishes.

If the assumption on the continuity of the transition used to derive
$p_{c}^{-}$ holds, the evolution equation around $p_{c}^{-}$ will also
show a continuous critical behavior at the value of $p_{c}^{-}$
obtained from Eq.~(\ref{eq7}). We solve numerically the
Eqs.~(\ref{eq2})-(\ref{eq3}) for $p_{c}^{-}+\delta p$, for different
values of $\alpha$. In the inset of Fig.~\ref{fig_target}a we show the
temporal evolution for $\alpha=0.01\%$ (with $p_{c}^{-}=0.077821334$).
We can see that above but very close to our theoretical $p_{c}^{-}$,
$\Psi_{\infty}$ goes to a finite value, while slightly below network
$A$ collapses ($\Psi_{\infty}=0$).  In Fig.~\ref{fig.phase} we show
(with cross symbols) some values of $p_{c}^{-}$ of the continuous
branch of the phase diagram obtained from the evolution
equations~\footnote{In order to determine $p_{c}^{-}$, we use $\delta
  p=10^{-12}$}, that are in total agreement with Eq.~(\ref{eq7}). This
result confirms our argument which leads to Eq.~(\ref{eq7}), used to
obtain the lower branch of the phase diagram allowing us to find
$p_{c}^{-}$. We found the same qualitative behavior for different
values of $\lambda$, however as the heterogeneity decreases, the
network is less robust and it is expected that at some point the
triple point will be lost. At this point the phase diagram will have
only one transition line, such as in homogeneous
networks~\cite{Sch_01}. These findings may indicate that when high
degree nodes in SF networks are protected via targeted autonomization,
random autonomization~\cite{Di_01} or correlation~\cite{Val_01}, they
induce multiple and different kind of order transitions.

In summary, we have presented the exact formulation of the cascade of
failures for targeted immunization with any degree distribution of
connectivity links. We show theoretically that increasing
autonomization $\alpha$ enhances the robustness of SF networks and
generates in the phase diagram $p-\alpha$ different regimes with
different characteristic sizes of the GC. These regimes converge into
a triple point which is a reminiscent of the triple points of
liquids. Physically it means that high degree nodes, that are
responsible to maintain the integrity of the networks, play a
fundamental role in the rich phase diagrams of these processes.
\subsection{Acknowledgments}
 L.D.V, P.A.M and L.A.B thank UNMdP and FONCyT (Pict 0293/2008) for
 financial support.

\bibliography{bib}

\begin{thebibliography}{27}
\expandafter\ifx\csname natexlab\endcsname\relax\def\natexlab#1{#1}\fi
\expandafter\ifx\csname bibnamefont\endcsname\relax
  \def\bibnamefont#1{#1}\fi
\expandafter\ifx\csname bibfnamefont\endcsname\relax
  \def\bibfnamefont#1{#1}\fi
\expandafter\ifx\csname citenamefont\endcsname\relax
  \def\citenamefont#1{#1}\fi
\expandafter\ifx\csname url\endcsname\relax
  \def\url#1{\texttt{#1}}\fi
\expandafter\ifx\csname urlprefix\endcsname\relax\def\urlprefix{URL }\fi
\providecommand{\bibinfo}[2]{#2}
\providecommand{\eprint}[2][]{\url{#2}}

\bibitem[{\citenamefont{Rinaldi et~al.}(2001)\citenamefont{Rinaldi, Peerenboom,
  and Kelly}}]{Rin_01}
\bibinfo{author}{\bibfnamefont{S.~M.} \bibnamefont{Rinaldi}},
  \bibinfo{author}{\bibfnamefont{J.~P.} \bibnamefont{Peerenboom}},
  \bibnamefont{and} \bibinfo{author}{\bibfnamefont{T.~K.} \bibnamefont{Kelly}},
  \bibinfo{journal}{Control Systems, IEEE} \textbf{\bibinfo{volume}{21}},
  \bibinfo{pages}{11} (\bibinfo{year}{2001}).

\bibitem[{\citenamefont{Kivel{\"a} et~al.}(2013)\citenamefont{Kivel{\"a},
  Arenas, Barthelemy, Gleeson, Moreno, and Porter}}]{Kiv_01}
\bibinfo{author}{\bibfnamefont{M.}~\bibnamefont{Kivel{\"a}}},
  \bibinfo{author}{\bibfnamefont{A.}~\bibnamefont{Arenas}},
  \bibinfo{author}{\bibfnamefont{M.}~\bibnamefont{Barthelemy}},
  \bibinfo{author}{\bibfnamefont{J.~P.} \bibnamefont{Gleeson}},
  \bibinfo{author}{\bibfnamefont{Y.}~\bibnamefont{Moreno}}, \bibnamefont{and}
  \bibinfo{author}{\bibfnamefont{M.~A.} \bibnamefont{Porter}},
  \bibinfo{journal}{arXiv preprint arXiv:1309.7233}  (\bibinfo{year}{2013}).

\bibitem[{\citenamefont{Rosato et~al.}(2008)\citenamefont{Rosato, Issacharoff,
  Tiriticco, Meloni, Porcellinis, and Setola}}]{Ros_01}
\bibinfo{author}{\bibfnamefont{V.}~\bibnamefont{Rosato}},
  \bibinfo{author}{\bibfnamefont{L.}~\bibnamefont{Issacharoff}},
  \bibinfo{author}{\bibfnamefont{F.}~\bibnamefont{Tiriticco}},
  \bibinfo{author}{\bibfnamefont{S.}~\bibnamefont{Meloni}},
  \bibinfo{author}{\bibfnamefont{S.}~\bibnamefont{Porcellinis}},
  \bibnamefont{and} \bibinfo{author}{\bibfnamefont{R.}~\bibnamefont{Setola}},
  \bibinfo{journal}{International Journal of Critical Infrastructures}
  \textbf{\bibinfo{volume}{4}}, \bibinfo{pages}{63} (\bibinfo{year}{2008}).

\bibitem[{\citenamefont{Parshani
  et~al.}(2010{\natexlab{a}})\citenamefont{Parshani, Rozenblat, Ietri, Ducruet,
  and Havlin}}]{Ron_02}
\bibinfo{author}{\bibfnamefont{R.}~\bibnamefont{Parshani}},
  \bibinfo{author}{\bibfnamefont{C.}~\bibnamefont{Rozenblat}},
  \bibinfo{author}{\bibfnamefont{D.}~\bibnamefont{Ietri}},
  \bibinfo{author}{\bibfnamefont{C.}~\bibnamefont{Ducruet}}, \bibnamefont{and}
  \bibinfo{author}{\bibfnamefont{S.}~\bibnamefont{Havlin}},
  \bibinfo{journal}{Europhys. Lett.} \textbf{\bibinfo{volume}{92}},
  \bibinfo{pages}{68002} (\bibinfo{year}{2010}{\natexlab{a}}).

\bibitem[{\citenamefont{Buldyrev et~al.}(2010)\citenamefont{Buldyrev, Parshani,
  Paul, Stanley, and Havlin}}]{Bul_02}
\bibinfo{author}{\bibfnamefont{S.~V.} \bibnamefont{Buldyrev}},
  \bibinfo{author}{\bibfnamefont{R.}~\bibnamefont{Parshani}},
  \bibinfo{author}{\bibfnamefont{G.}~\bibnamefont{Paul}},
  \bibinfo{author}{\bibfnamefont{H.~E.} \bibnamefont{Stanley}},
  \bibnamefont{and} \bibinfo{author}{\bibfnamefont{S.}~\bibnamefont{Havlin}},
  \bibinfo{journal}{Nature} \textbf{\bibinfo{volume}{464}},
  \bibinfo{pages}{1025} (\bibinfo{year}{2010}).

\bibitem[{\citenamefont{Parshani
  et~al.}(2010{\natexlab{b}})\citenamefont{Parshani, Buldyrev, and
  Havlin}}]{Ron_01}
\bibinfo{author}{\bibfnamefont{R.}~\bibnamefont{Parshani}},
  \bibinfo{author}{\bibfnamefont{S.~V.} \bibnamefont{Buldyrev}},
  \bibnamefont{and} \bibinfo{author}{\bibfnamefont{S.}~\bibnamefont{Havlin}},
  \bibinfo{journal}{Phys. Rev. Lett.} \textbf{\bibinfo{volume}{105}},
  \bibinfo{pages}{048701} (\bibinfo{year}{2010}{\natexlab{b}}).

\bibitem[{\citenamefont{Zhou et~al.}(2013)\citenamefont{Zhou, Gao, Stanley, and
  Havlin}}]{Di_01}
\bibinfo{author}{\bibfnamefont{D.}~\bibnamefont{Zhou}},
  \bibinfo{author}{\bibfnamefont{J.}~\bibnamefont{Gao}},
  \bibinfo{author}{\bibfnamefont{H.~E.} \bibnamefont{Stanley}},
  \bibnamefont{and} \bibinfo{author}{\bibfnamefont{S.}~\bibnamefont{Havlin}},
  \bibinfo{journal}{Phys. Rev. E} \textbf{\bibinfo{volume}{87}},
  \bibinfo{pages}{052812} (\bibinfo{year}{2013}).

\bibitem[{\citenamefont{Gao et~al.}(2011{\natexlab{a}})\citenamefont{Gao,
  Buldyrev, Havlin, and Stanley}}]{Gao_03}
\bibinfo{author}{\bibfnamefont{J.}~\bibnamefont{Gao}},
  \bibinfo{author}{\bibfnamefont{S.~V.} \bibnamefont{Buldyrev}},
  \bibinfo{author}{\bibfnamefont{S.}~\bibnamefont{Havlin}}, \bibnamefont{and}
  \bibinfo{author}{\bibfnamefont{H.~E.} \bibnamefont{Stanley}},
  \bibinfo{journal}{Phys. Rev. Lett.} \textbf{\bibinfo{volume}{107}},
  \bibinfo{pages}{195701} (\bibinfo{year}{2011}{\natexlab{a}}).

\bibitem[{\citenamefont{Gao et~al.}(2011{\natexlab{b}})\citenamefont{Gao,
  Buldyrev, Stanley, and Havlin}}]{Gao_01}
\bibinfo{author}{\bibfnamefont{J.}~\bibnamefont{Gao}},
  \bibinfo{author}{\bibfnamefont{S.~V.} \bibnamefont{Buldyrev}},
  \bibinfo{author}{\bibfnamefont{H.~E.} \bibnamefont{Stanley}},
  \bibnamefont{and} \bibinfo{author}{\bibfnamefont{S.}~\bibnamefont{Havlin}},
  \bibinfo{journal}{Nature Physics} \textbf{\bibinfo{volume}{8}},
  \bibinfo{pages}{40} (\bibinfo{year}{2011}{\natexlab{b}}).

\bibitem[{\citenamefont{Gao et~al.}(2013)\citenamefont{Gao, Buldyrev, Stanley,
  Xu, and Havlin}}]{Gao_02}
\bibinfo{author}{\bibfnamefont{J.}~\bibnamefont{Gao}},
  \bibinfo{author}{\bibfnamefont{S.~V.} \bibnamefont{Buldyrev}},
  \bibinfo{author}{\bibfnamefont{H.~E.} \bibnamefont{Stanley}},
  \bibinfo{author}{\bibfnamefont{X.}~\bibnamefont{Xu}}, \bibnamefont{and}
  \bibinfo{author}{\bibfnamefont{S.}~\bibnamefont{Havlin}},
  \bibinfo{journal}{arXiv preprint arXiv:1306.3416}  (\bibinfo{year}{2013}).

\bibitem[{Exx({\natexlab{a}})}]{Exxon_02}
\bibinfo{note}{Note that since in a SF network there is only a few amount of
  high degree nodes, they have a low probability to become dysfunctional in the
  initial failure.}

\bibitem[{\citenamefont{Schneider et~al.}(2013)\citenamefont{Schneider,
  Yazdani, Ara{\'u}jo, Havlin, and Herrmann}}]{Sch_01}
\bibinfo{author}{\bibfnamefont{C.~M.} \bibnamefont{Schneider}},
  \bibinfo{author}{\bibfnamefont{N.}~\bibnamefont{Yazdani}},
  \bibinfo{author}{\bibfnamefont{N.~A.} \bibnamefont{Ara{\'u}jo}},
  \bibinfo{author}{\bibfnamefont{S.}~\bibnamefont{Havlin}}, \bibnamefont{and}
  \bibinfo{author}{\bibfnamefont{H.~J.} \bibnamefont{Herrmann}},
  \bibinfo{journal}{Scientific reports} \textbf{\bibinfo{volume}{3}},
  \bibinfo{pages}{1969} (\bibinfo{year}{2013}).

\bibitem[{\citenamefont{Yagan et~al.}(2012)\citenamefont{Yagan, Qian, Zhang,
  and Cochran}}]{Yag_01}
\bibinfo{author}{\bibfnamefont{O.}~\bibnamefont{Yagan}},
  \bibinfo{author}{\bibfnamefont{D.}~\bibnamefont{Qian}},
  \bibinfo{author}{\bibfnamefont{J.}~\bibnamefont{Zhang}}, \bibnamefont{and}
  \bibinfo{author}{\bibfnamefont{D.}~\bibnamefont{Cochran}},
  \bibinfo{journal}{Parallel and Distributed Systems, IEEE Transactions on}
  \textbf{\bibinfo{volume}{23}}, \bibinfo{pages}{1708} (\bibinfo{year}{2012}).

\bibitem[{\citenamefont{Valdez et~al.}(2013)\citenamefont{Valdez, Macri,
  Stanley, and Braunstein}}]{Val_01}
\bibinfo{author}{\bibfnamefont{L.}~\bibnamefont{Valdez}},
  \bibinfo{author}{\bibfnamefont{P.}~\bibnamefont{Macri}},
  \bibinfo{author}{\bibfnamefont{H.}~\bibnamefont{Stanley}}, \bibnamefont{and}
  \bibinfo{author}{\bibfnamefont{L.}~\bibnamefont{Braunstein}},
  \bibinfo{journal}{Phys. Rev. E} \textbf{\bibinfo{volume}{88}},
  \bibinfo{pages}{050803} (\bibinfo{year}{2013}).

\bibitem[{\citenamefont{Braunstein et~al.}(2007)\citenamefont{Braunstein, Wu,
  Chen, Buldyrev, Kalisky, Sreenivasan, Cohen, L{\'o}pez, Havlin, and
  Stanley}}]{Bra_01}
\bibinfo{author}{\bibfnamefont{L.~A.} \bibnamefont{Braunstein}},
  \bibinfo{author}{\bibfnamefont{Z.}~\bibnamefont{Wu}},
  \bibinfo{author}{\bibfnamefont{Y.}~\bibnamefont{Chen}},
  \bibinfo{author}{\bibfnamefont{S.~V.} \bibnamefont{Buldyrev}},
  \bibinfo{author}{\bibfnamefont{T.}~\bibnamefont{Kalisky}},
  \bibinfo{author}{\bibfnamefont{S.}~\bibnamefont{Sreenivasan}},
  \bibinfo{author}{\bibfnamefont{R.}~\bibnamefont{Cohen}},
  \bibinfo{author}{\bibfnamefont{E.}~\bibnamefont{L{\'o}pez}},
  \bibinfo{author}{\bibfnamefont{S.}~\bibnamefont{Havlin}}, \bibnamefont{and}
  \bibinfo{author}{\bibfnamefont{H.~E.} \bibnamefont{Stanley}},
  \bibinfo{journal}{International Journal of Bifurcation and Chaos}
  \textbf{\bibinfo{volume}{17}}, \bibinfo{pages}{2215} (\bibinfo{year}{2007}).

\bibitem[{\citenamefont{Newman et~al.}(2001)\citenamefont{Newman, Strogatz, and
  Watts}}]{New_01}
\bibinfo{author}{\bibfnamefont{M.}~\bibnamefont{Newman}},
  \bibinfo{author}{\bibfnamefont{S.}~\bibnamefont{Strogatz}}, \bibnamefont{and}
  \bibinfo{author}{\bibfnamefont{D.}~\bibnamefont{Watts}},
  \bibinfo{journal}{Phys. Rev. E} \textbf{\bibinfo{volume}{64}},
  \bibinfo{pages}{026118} (\bibinfo{year}{2001}).

\bibitem[{\citenamefont{Son et~al.}(2012)\citenamefont{Son, Bizhani,
  Christensen, Grassberger, and Paczuski}}]{Son_01}
\bibinfo{author}{\bibfnamefont{S.-W.} \bibnamefont{Son}},
  \bibinfo{author}{\bibfnamefont{G.}~\bibnamefont{Bizhani}},
  \bibinfo{author}{\bibfnamefont{C.}~\bibnamefont{Christensen}},
  \bibinfo{author}{\bibfnamefont{P.}~\bibnamefont{Grassberger}},
  \bibnamefont{and} \bibinfo{author}{\bibfnamefont{M.}~\bibnamefont{Paczuski}},
  \bibinfo{journal}{Europhys. Lett.} \textbf{\bibinfo{volume}{97}},
  \bibinfo{pages}{16006} (\bibinfo{year}{2012}).

\bibitem[{\citenamefont{Baxter et~al.}(2012)\citenamefont{Baxter, Dorogovtsev,
  Goltsev, and Mendes}}]{Bax_01}
\bibinfo{author}{\bibfnamefont{G.}~\bibnamefont{Baxter}},
  \bibinfo{author}{\bibfnamefont{S.}~\bibnamefont{Dorogovtsev}},
  \bibinfo{author}{\bibfnamefont{A.}~\bibnamefont{Goltsev}}, \bibnamefont{and}
  \bibinfo{author}{\bibfnamefont{J.}~\bibnamefont{Mendes}},
  \bibinfo{journal}{Phys. Rev. Lett.} \textbf{\bibinfo{volume}{109}},
  \bibinfo{pages}{248701} (\bibinfo{year}{2012}).

\bibitem[{\citenamefont{Bornholdt et~al.}(2003)\citenamefont{Bornholdt,
  Schuster, and Wiley}}]{Bor_01}
\bibinfo{author}{\bibfnamefont{S.}~\bibnamefont{Bornholdt}},
  \bibinfo{author}{\bibfnamefont{H.~G.} \bibnamefont{Schuster}},
  \bibnamefont{and} \bibinfo{author}{\bibfnamefont{J.}~\bibnamefont{Wiley}},
  \emph{\bibinfo{title}{Handbook of graphs and networks}},
  vol.~\bibinfo{volume}{2} (\bibinfo{publisher}{Wiley Online Library},
  \bibinfo{year}{2003}).

\bibitem[{\citenamefont{Nikolaevskaya et~al.}(2012)\citenamefont{Nikolaevskaya,
  Khimich, and Chistyakova}}]{Nik_01}
\bibinfo{author}{\bibfnamefont{E.}~\bibnamefont{Nikolaevskaya}},
  \bibinfo{author}{\bibfnamefont{A.}~\bibnamefont{Khimich}}, \bibnamefont{and}
  \bibinfo{author}{\bibfnamefont{T.}~\bibnamefont{Chistyakova}},
  \emph{\bibinfo{title}{Programming with multiple precision}}, vol.
  \bibinfo{volume}{397} (\bibinfo{publisher}{Springer}, \bibinfo{year}{2012}).

\bibitem[{\citenamefont{Amaral et~al.}(2000)\citenamefont{Amaral, Scala,
  Barth{\'e}l{\'e}my, and Stanley}}]{Ama_01}
\bibinfo{author}{\bibfnamefont{L.~A.~N.} \bibnamefont{Amaral}},
  \bibinfo{author}{\bibfnamefont{A.}~\bibnamefont{Scala}},
  \bibinfo{author}{\bibfnamefont{M.}~\bibnamefont{Barth{\'e}l{\'e}my}},
  \bibnamefont{and} \bibinfo{author}{\bibfnamefont{H.~E.}
  \bibnamefont{Stanley}}, \bibinfo{journal}{Proceedings of the National Academy
  of Sciences} \textbf{\bibinfo{volume}{97}}, \bibinfo{pages}{11149}
  (\bibinfo{year}{2000}).

\bibitem[{\citenamefont{Barab{\'a}si and Albert}(1999)}]{Bar_02}
\bibinfo{author}{\bibfnamefont{A.-L.} \bibnamefont{Barab{\'a}si}}
  \bibnamefont{and} \bibinfo{author}{\bibfnamefont{R.}~\bibnamefont{Albert}},
  \bibinfo{journal}{Science} \textbf{\bibinfo{volume}{286}},
  \bibinfo{pages}{509} (\bibinfo{year}{1999}).

\bibitem[{\citenamefont{Barab{\'a}si et~al.}(2002)\citenamefont{Barab{\'a}si,
  Jeong, N{\'e}da, Ravasz, Schubert, and Vicsek}}]{Bar_01}
\bibinfo{author}{\bibfnamefont{A.-L.} \bibnamefont{Barab{\'a}si}},
  \bibinfo{author}{\bibfnamefont{H.}~\bibnamefont{Jeong}},
  \bibinfo{author}{\bibfnamefont{Z.}~\bibnamefont{N{\'e}da}},
  \bibinfo{author}{\bibfnamefont{E.}~\bibnamefont{Ravasz}},
  \bibinfo{author}{\bibfnamefont{A.}~\bibnamefont{Schubert}}, \bibnamefont{and}
  \bibinfo{author}{\bibfnamefont{T.}~\bibnamefont{Vicsek}},
  \bibinfo{journal}{Physica A: Statistical Mechanics and its Applications}
  \textbf{\bibinfo{volume}{311}}, \bibinfo{pages}{590} (\bibinfo{year}{2002}).

\bibitem[{\citenamefont{Jeong et~al.}(2001)\citenamefont{Jeong, Mason,
  Barab{\'a}si, and Oltvai}}]{Jeo_01}
\bibinfo{author}{\bibfnamefont{H.}~\bibnamefont{Jeong}},
  \bibinfo{author}{\bibfnamefont{S.~P.} \bibnamefont{Mason}},
  \bibinfo{author}{\bibfnamefont{A.-L.} \bibnamefont{Barab{\'a}si}},
  \bibnamefont{and} \bibinfo{author}{\bibfnamefont{Z.~N.}
  \bibnamefont{Oltvai}}, \bibinfo{journal}{Nature}
  \textbf{\bibinfo{volume}{411}}, \bibinfo{pages}{41} (\bibinfo{year}{2001}).

\bibitem[{Exx({\natexlab{b}})}]{Exxon_03}
\bibinfo{note}{In Ref.~\cite{Sch_01} it was found only one threshold, since the
  process was studied only on homogeneous networks.}

\bibitem[{Exx({\natexlab{c}})}]{Exxon_01}
\bibinfo{note}{Geometrically, this equation is the condition of the tangency
  between the identity plane and the surface composed by the right hand side of
  Eqs.~(\ref{eq2}) and (\ref{eq4}) in the steady state.}

\bibitem[{Exx({\natexlab{d}})}]{Exxon_04}
\bibinfo{note}{It is straightforward to obtain the same result for random
  autonomization~\cite{Di_01} under random failure in both networks.}

\end{thebibliography}
\end{document}